# Spectral mapping of polarization-correlated photon-pair sources using quantum-classical correspondence

**HUNG-PIN CHUNG,**[1,†] **PAWAN KUMAR,**[2,†] **KAI WANG,**[3,4] **OLIVIER BERNARD,**[2,5] **CHINMAY SHIRPURKAR,**[2,6] **WEN-CHIUAN SU,**[1] **THOMAS PERTSCH**[2], **ANDREY A. SUKHORUKOV,**[3] **YEN-HUNG CHEN,**[1,7,*] **AND FRANK SETZPFANDT**[2,*]

[1]*Department of Optics and Photonics, National Central University, Jhongli 320, Taiwan*
[2]*Institute of Applied Physics, Abbe Center of Photonics, Friedrich-Schiller-Universität Jena, 07745 Jena, Germany*
[3]*Nonlinear Physics Centre, Research School of Physics, The Australian National University, Canberra, ACT 2601, Australia*
[4]*Ginzton Laboratory and Department of Electrical Engineering, Stanford University, Stanford, CA 94305, USA*
[5]*Galatea Laboratory, STI/IMT, École polytechnique fédérale de Lausanne (EPFL), 2002 Neuchâtel, Switzerland*
[6]*CREOL, College of Optics and Photonics, University of Central Florida, Orlando, FL 32816, USA*
[7]*Center for Astronautical Physics and Engineering, National Central University, Jhongli 320, Taiwan*
[†] *These authors contributed equally*

*[*]yhchen@dop.ncu.edu.tw, f.setzpfandt@uni-jena.de*

**Abstract:** Direct spectral characterization of a quantum photon-pair source usually involves cumbersome, costly, and time-consuming detection issues. In this study, we experimentally characterize the spectral properties of a type-II phase-matched spontaneous parametric down-conversion (SPDC) source based on a titanium-diffused periodically poled lithium niobate (Ti:PPLN) waveguide. The characterization of the spectral information of the generated cross-polarized photon pairs is of importance for the use of such sources in applications including quantum information and communication. We demonstrate that the joint spectral intensity of the cross-polarized photon-pair source can be fully reconstructed using the quantum-classical correspondence through classical sum-frequency generation (SFG) measurements. This technique, which uses a much less complex detection system for visible light, opens the possibility of fast monitoring and control of the quantum state of (polarization-correlated) photon-pair sources to facilitate the realization of a stable and high-usability quantum source.

## 1. Introduction

In the past decades, quantum optics significantly progressed, from the illustrious experimental tests of Bell's inequalities and quantum teleportation [1, 2] to practical demonstrations of applications such as the quantum key distribution (QKD) and towards realizing unconditionally secure communications [3, 4]. To realize such involved quantum applications, the implementation of optimized sources of photonic quantum states and their precise characterization is of utmost importance. One of the most prominent mechanisms to generate quantum light is spontaneous parametric down-conversion (SPDC) in nonlinear media with second-order (or quadratic) nonlinearity. Here, a propagating photon at the pump frequency $\omega_p$ is split into a pair of signal and idler photons with frequencies $\omega_s$ and $\omega_i$, respectively. Due to energy conservation the frequencies of the three photons obey $\omega_p = \omega_s + \omega_i$. The photon pairs probabilistically generated by this process can be used to either realize heralded single-photons, where the idler photon is used to herald the presence of the signal photon, or entangled pairs, where the properties of the nonlinear medium determine the degree of freedom in which entanglement appears as well as its amount. Typical nonlinear crystals for photon-pair generation are potassium titanyl phosphate (KTP), barium borate (BBO), or lithium niobate ($LiNbO_3$). For efficient photon-pair generation, phase matching between the signal, idler, and pump waves has to be achieved. This can be done by adjusting the crystal temperature, the propagation direction in birefringent crystals, or by using quasi-phase-matching (QPM), where the crystal structure is periodically inverted to create additional wavevectors for phase matching.

To enhance the generation efficiency, SPDC can be implemented in waveguides [5], where light can be much more confined compared to bulk crystals to effectively increase the light intensity and the interaction length and therefore the nonlinear conversion efficiency. In $LiNbO_3$, waveguides with QPM can be realized with lengths of several cm, leading to strongly increased source brightness and enhanced stability in contrast to their bulk counterparts.

Of particular interest are photon-pair sources using the so-called type-II phase matching, where the generated signal and idler photons are in orthogonal polarizations. Such sources can be used to create photon pairs entangled in polarization [6], which have been utilized in many demonstrations of fundamental quantum effects and in quantum information processing, including photon interference at nano-resonators and imaging with metasurfaces [7, 8]. Furthermore, type-II SPDC is also very useful for heralded single photon generation, as the signal and idler photons can be easily separated based on their polarization.

For any of these applications of type-II SPDC, control of the joint-spectral amplitude (JSA) of the generated photons is needed. To generate entanglement, two SPDC processes should be superposed, where the photons from either process need to be spectrally indistinguishable. To generate pure single photons, no spectral entanglement should be present, i.e. the JSA should be factorable. Finally, for applications, the spectral properties of the photon-pair source need to be adapted to the other parts of the used optical system, e.g. to the spectral lines of the ITU grid for telecom applications. This necessitates a specific design of the sources to adapt the spectral properties to applications, but also calls for fast and robust spectral characterization methods for the qualification of fabricated sources [9].

For direct measurement of the JSA of a photon-pair source, spectrally resolved correlation measurements have to be performed. This can be achieved by spectrally filtering e.g. the signal photon and measuring the idler photon spectrum, typically by scanning a monochromator, conditional on signal detections [10, 11]. This process has to be repeated for any signal wavelength of interest to reconstruct the complete JSA. Due to the needed spectral filtering, typically only a small number of photons will be available for any given measurement, leading to long integration times. This is especially true when type-II SPDC sources need to be characterized. Here, the relatively large difference in group velocity between the two down-converted polarization modes leads to SPDC with much narrower spectra in contrast to other SPDC schemes. Hence, very narrow spectral filtering is needed to obtain sufficient information about the JSA.

An alternative approach for fast characterization of the JSA is based on the stimulated emission of the down-converted photons using difference-frequency generation [12]. Here, the number of emitted photons will be proportional to that of the spontaneous process with a proportionality constant depending on the flux of the stimulating laser [13-18]. Due to this, the number of the generated classical photons in the stimulated process can be many orders of magnitude higher than that from the spontaneous process. Hence, the stimulated-emission based measurement can be much more efficient without the use of single-photon detectors, leading to an improved signal-to-noise ratio. This method, however, is not suitable for applications in characterizing quantum networks built in integrated optical circuits where multimode nonlinear interaction with finite and possibly, various types of losses are involved [19].

Recent studies showed the issue of losses in real devices can be readily overcome with the use of classical sum-frequency generation (SFG), the reverse process of difference frequency generation [19]. This has the added advantage, that for the most common SPDC sources which pump laser wavelengths are in the near-visible region, detectors based on silicon can be used, further improving the signal-to-noise ratio of the measurement. More recently, this SFG-SPDC analogy technique has been formulated and demonstrated for lossy multimode nonlinear quantum photonic circuits [20] and nano-resonators [21], based on a general analytical solution for SPDC [22]. In Ref. [20], the SFG-SPDC analogy has been used to characterize a type-0 PPLN waveguide circuit supporting only TM polarization modes.

In this study, we experimentally demonstrate the feasibility of the SFG-SPDC analogy for joint spectral characterization of cross-polarized photon pairs generated in a type-II PPLN SPDC waveguide source. By utilizing the quantum-classical correspondence relationship in the SFG-SPDC spectral mapping [20], the full spectral information of the nonclassical source is acquired. In this way, we experimentally verify the key features of a type-II SPDC source, namely a narrow spectrum even close to degenerate signal and idler wavelengths and the absence of an upper limit for the usable pump wavelengths to reach phase-matching as in type-0 SPDC [23]. The experiments on quantum-classical spectral coincidence mapping in this work suggest a feasible solution allowing to quickly monitor and control the quantum states of photon-pair sources even with narrowband polarization-entangled characteristics via some dynamic feedback mechanism based on e.g. temperature or/and pump wavelength tuning compensations [24] to facilitate the stable production of heralded single photons or entangled photon pairs with the desired degree of purity and indistinguishability.

## 2. Characterization of polarization-correlated photon-pair source

*2.1 Preparation of a type-II SPDC quantum source*

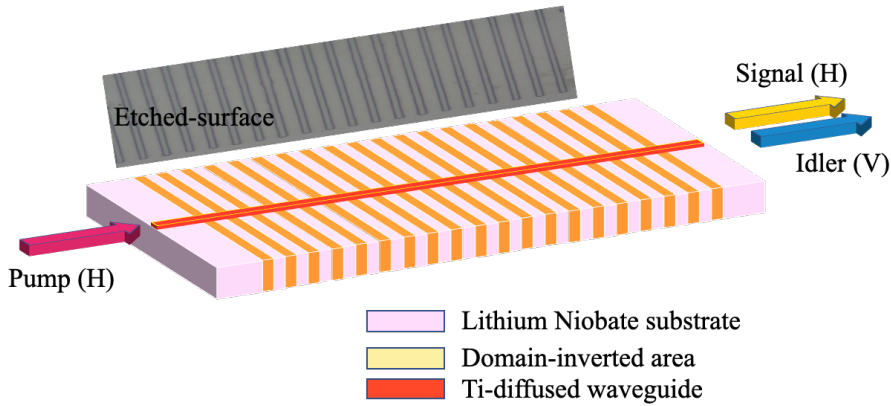

Fig. 1. Schematic of a type-II Ti:PPLN SPDC waveguide chip. The inset shows a typical microscopic image of a portion of the +z surface of the etched PPLN sample.

The photon-pair source used in this study was based on z-cut titanium-indiffused periodically poled lithium niobate (Ti:PPLN) waveguide chips. For generating cross-polarized signal and idler photon pairs, a type-II QPM SPDC process is employed, where a cw TE (H)-polarized laser tunable around 770 nm is used as the pump. Fig. 1 shows a sketch of such a device. The waveguides were fabricated using the titanium thermal diffusion (TTD) method [25] and the QPM domain structure was formed by using the standard electric field poling technique [26]; details on the device fabrication methodology can be found elsewhere [27]. The waveguides support low-loss guiding of the two cross-polarized fundamental modes in the 1450-1650 nm spectral region with typical losses of ~0.3 and ~0.6 dB/cm for V-polarized idler and H-polarized signal modes, respectively. In consideration of a possible fabrication error, we have realized 9 PPLN SPDC waveguides along the crystallographic y axis in a 20-mm long, 10-mm wide, and

0.5 mm thick LiNbO$_3$ chip with different QPM periods from 8.9 to 9.7 μm with an increment of 0.1 μm between different waveguides. The waveguides and the QPM are designed to operate at temperatures larger than 120° C to avoid the adverse effect from photorefraction possibly induced in the crystal by the near-visible pump laser.

First, the fabricated sample was characterized classically using degenerate SFG, where the exciting signal and idler waves had the same wavelengths. To this end, a cw laser tunable in the telecom 1.5 μm band was coupled to the waveguide after passing a 45° polarizer to simultaneously excite the V-polarized and H-polarized modes for performing SFG. Fig. 2(a) shows the measured phase-matching wavelengths as a function of the PPLN grating period for various operating temperatures. These phase-matching results correspond to the wavelengths where SPDC with degenerate signal and idler is expected. The maximal conversion efficiency of the waveguide is approximately 2 %W$^{-1}$. Fig. 2(b) shows the characterized phase-matching spectra of the waveguide for type-II degenerate SFG with a pump power of 178 μW for two different sample temperatures. The results indicate a narrow phase-matching bandwidth of ~0.43 nm. This bandwidth is smaller but of the same order of magnitude compared to a type-0 SFG process. However, in SPDC, which has much lower generation efficiency compared to SFG, type-II sources have a much smaller bandwidth due to the larger difference in the group velocity of signal and idler compared to type-0 SPDC. Hence, they demand a stricter control of the operating conditions, such as the pump wavelength and crystal temperature.

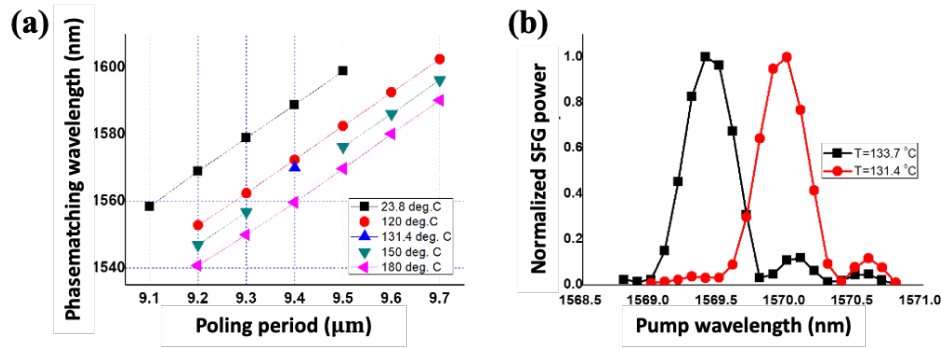

Fig. 2. Results measured with degenerate SFG. (a) Phase-matching wavelengths as a function of the PPLN grating period for various operating temperatures. (b) Temperature tuned phase-matching spectra at a pump power of 178 μW.

## 2.2 Classical SFG spectral characterization

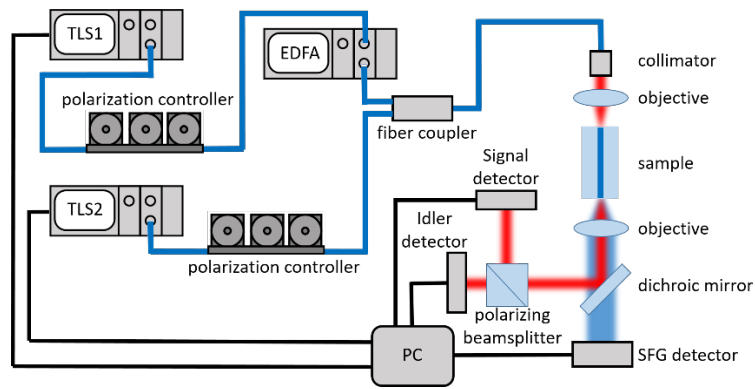

Fig. 3. Scheme of the experimental setup used for the classical characterization of the joint spectral properties from the SPDC waveguide sample using SFG.

Next, we demonstrate the joint spectral characterization of a type-II SPDC source based on SFG according to the quantum-classical correspondence theory. In this proof-of-principle experiment, a Ti:PPLN waveguide source with 9.4-μm poling period was used. The sample had a shorter length of 10 mm than the one used for initial testing to increase the type-II SPDC bandwidth to theoretically estimated 3.68 nm. This ensures that we can corroborate the SFG measurements with SPDC measurements using a spectrometer of limited resolution (see section 2.3). The setup to measure the SFG characteristics of the SPDC waveguide source is depicted in Fig. 3. Two independent tunable laser sources were used with their polarizations adjusted to horizontal and vertical, respectively, using fiber-based polarization controllers. The wavelengths of the two fundamental lasers are independently scanned in the telecom C band (~1530-1565 nm). To increase the excitation power, a polarization-maintaining erbium-doped fiber amplifier furthermore amplified one of the input beams. The two lasers were combined by a fiber coupler and then coupled to the waveguide sample using a collimator and a microscope objective. Thus, the two cross-polarized lasers (called signal and idler beams, respectively, to be associated with the naming convention of the SPDC process) are simultaneously coupled to the waveguide to excite H- and V-polarized modes, respectively. The Ti:PPLN waveguide chip is heated to 220°C to avoid photorefraction. A dichroic mirror is used to separate the output SFG signal and the non-depleted fundamental waves, the cross-polarized signal and idler waves are further separated by using a polarizing beam splitter (PBS) for the respective power and spectral measurements. The scanning resolutions of the signal and idler lasers are 1 nm and 0.025 nm, their input powers are about 0.5 mW and 2.5 mW, respectively. Fig. 4(a) shows the intensity distribution of the SFG measured from the waveguide device over the tuning range of the two fundamental lasers, where the peak values fulfill the QPM conditions of the SFG signals around 771 nm. The calculated spectral dependence of the phase mismatch is plotted in Fig. 4(b), where the expected line of zero phase mismatch corresponds well to the measurement. The distribution shown in Fig. 4(a) represents the joint spectral information of the source from the SFG based measurement.

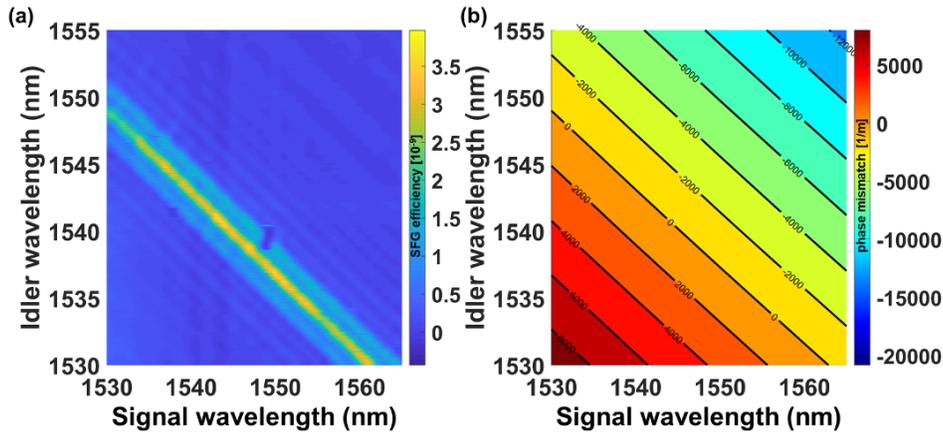

Fig. 4. (a) Intensity distribution of the SFG measured from the SPDC source as a function of the signal and idler wavelengths and (b) the corresponding numerical modelling of the phase-mismatching diagram over the tuning range of the two fundamental lasers.

*2.3 Direct SPDC spectral measurements*

To corroborate the SFG measurement results and to extend the measured spectral range, we further characterized the photon-pair spectra generated in the Ti:PPLN waveguide when pumped by a tunable cw laser in the wavelength range 765-780 nm by using a classical spectrometer. These measurements where performed using a setup separate from the one depicted in Fig. 3. Importantly, as SPDC and SFG are counter-directional processes [20], for the SPDC spectral measurements the sample was pumped from the side the generated SFG was emitted from in the experiments described earlier. However, we note that the used sample should be symmetric with respect to the propagation direction. The pump laser is linearly polarized and is aligned to excite a TE (*H*) polarization mode in the waveguide at a pump power of 22 mW. The cross-polarized SPDC photon-pair spectra were acquired by performing classical intensity measurement of the signal and idler waves via an IR spectrometer with a CCD line detector in the 1420-1690 nm range. The pump wavelength was changed in 1 nm steps and at each step, a spectrum was obtained with 200 s integration time. Fig. 5 shows the measured spectra in dependence on the pump wavelength, where signal and idler are not distinguished. Two crossed lines with prominent intensity, corresponding to the spectral positions of the two photons of the generated pairs, are observed. Spectral information at the crossing point, which corresponds to wavelength-degenerate SPDC, was not available due to the presence of strong residual pump radiation, measured despite the use of long pass filters before the detection system. The residual pump was measured in the 2nd grating order of the spectrometer and masked the weaker intensity of the photon pairs close to the degeneracy wavelength. We numerically removed those measured data in the corresponding spectral regions, marked by the dashed black lines in Fig. 5.

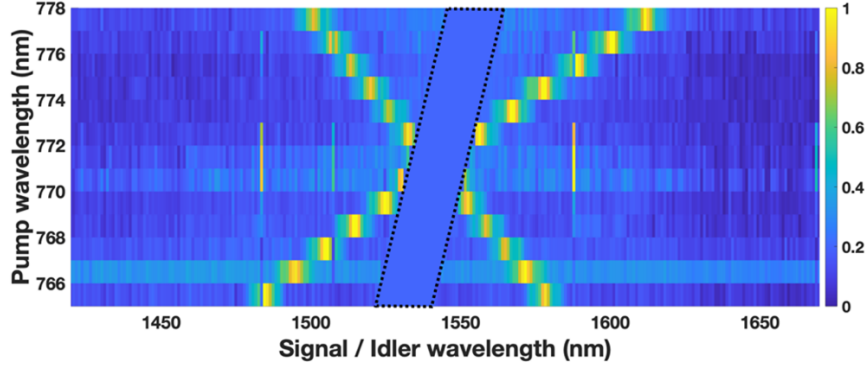

Fig. 5. Normalized measured SPDC spectra in dependence on the pump wavelength. Spectral information within the region marked by the dashed black lines is not available (see text).

## 3. Reconstruction of SPDC spectra using polarization-correlated SFG-SPDC analogy

*3.1 Coincidence mapping*

SFG can be regarded as the inverse process of SPDC in the stimulated emission regime. The correspondence between the bi-photon field states and the classically characterized SFG field has been studied [20], from which the photon-pair generation rate of a SPDC quantum process can be estimated based on the measured SFG information including the phase-matched wavelengths, spectral bandwidths, and conversion efficiencies, under the undepleted pump regime, given by

$$\frac{1}{P_p}\frac{dN_{pair}}{dt} = \sum_j \eta_j^{SFG} \frac{\lambda_p^2}{(\lambda_s)_j (\lambda_i)_j} \frac{c\Delta\lambda}{[(\lambda_s)_j]^2}, \qquad (1)$$

where $P_p$ is the pump power, $N_{pair}$ is the number of photon pairs, $\eta^{SFG}$ is the conversion efficiency, $\lambda_p$, $\lambda_s$, and $\lambda_i$ are wavelengths of pump, signal, idler waves, respectively, $\Delta\lambda$ is the wavelength scanning step (in signal or idler), and $c$ is the speed of light. We are using this quantum-classical correspondence relationship also to estimate the spectral dependence of the single photon counts of the SPDC in the near-degenerate region, which could not be measured using the classical spectrometer (marked region in Fig. 5), by utilizing those spectral data obtained in the SFG measurement (see Fig. 4(a)). Obtaining the absolute number of single counts using this method is strictly only possible in the case of lossless photonic circuits, where the number of single counts in the signal and idler is the same as $N_{pair}$. The losses in our waveguide for each involved spectral component are around 10% and are constant across the considered spectral region due to the low dispersion of the waveguide. In this case, the measured single-photon spectra contain contributions stemming from photon-pairs states and single-photon mixed states, where either the signal or the idler photon has been absorbed [28]. Using Eq. (1) for the calculation of $N_{pair}$ from the SFG intensity will only allow reconstruction of the photon-pair contribution to the SPDC spectra. However, in our case this still represents a reasonable estimation of the actual spectra.

To demonstrate this, in Fig. 6 we present and compare detailed spectra of several near-degenerate SPDC photon pairs obtained by the different methods. In Fig. 6(a), we show the signal-idler spectra as measured with the spectrometer. The peak values of all spectra are normalized; the uneven peak intensities in the spectra are due to different coupling efficiencies of the measurement system for the two polarizations. A theoretically calculated SPDC spectrum (dotted blue line) of a near-degenerate photon pair corresponding to 773 nm pump wavelength is also plotted in Fig. 6(a) for comparison. It shows that the measured spectra are in good agreement with the calculated ones in the peak regions (central lobes, with FWHM widths of ~3 nm) but show enhanced side lobes, which can be attributed to the axial variations in the waveguide width due to fabrication imperfections, resulting in axial phase velocity inhomogeneity [29].

In Fig. 6(b), we show spectra reconstructed from the SFG measurements corresponding to a pump wavelength of 772 nm, where the individually reconstructed signal (H-polarized) and idler (V-polarized) spectra are shown; and corresponding to a pump wavelength of 771 nm, close to the degeneracy, where we show only the H-polarized idler spectrum. The reconstructed spectra are also normalized to 1, the different number of data points for H and V polarization is due to the different wavelength steps used for the two lasers in the SFG experiment. The spectra obtained with both methods show similar features, i.e. bandwidths of around 3-4 nm FWHM in the central lobes and pronounced side lobes due to inhomogeneities. Hence, for our sample, approximate single-photon spectra can be obtained using the quantum-classical correspondence through Eq. (1).

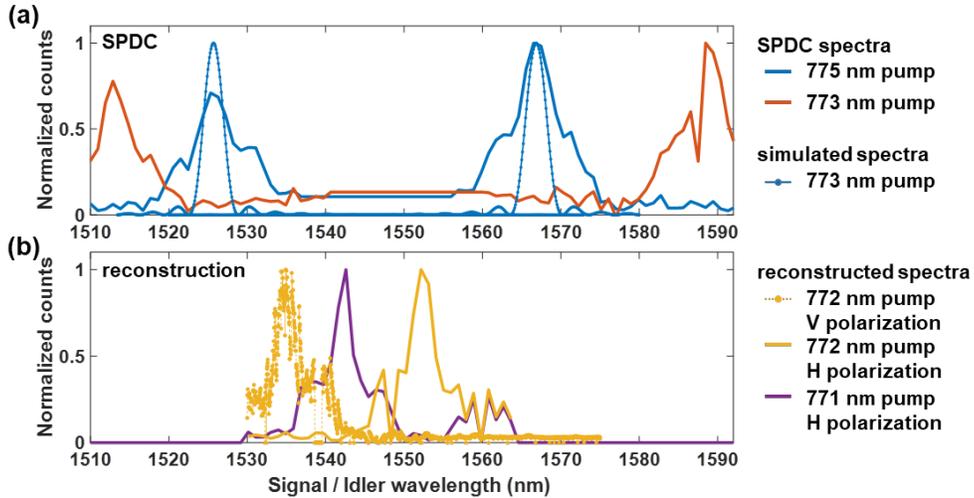

Fig. 6. Spectra of some near-degenerate SPDC photon pairs. (a) Directly measured (solid lines) and calculated (dotted line). (b) Reconstructed from the SFG measurements.

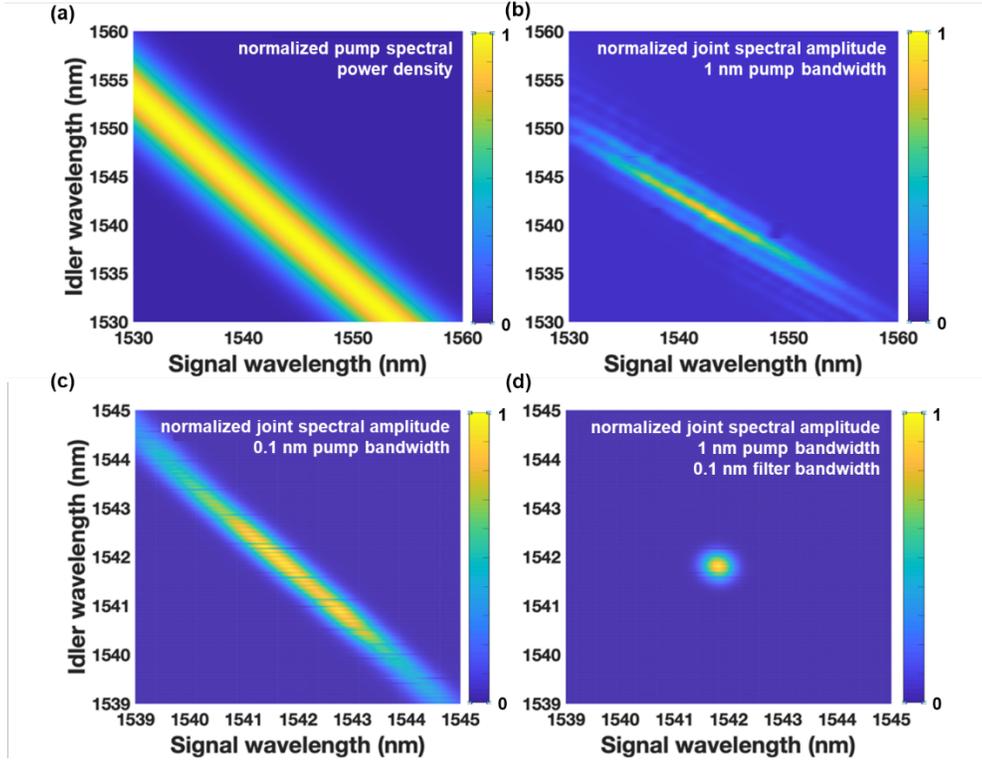

Fig. 7. (a) Joint spectral power density of pump beam with Gaussian spectrum centered at 770.9 nm and with 1 nm bandwidth. (b)-(d) JSAs of the type-II SPDC source near degeneracy, calculated from the classical SFG measurement when assuming a pump at 770.9 nm with (b) 1 nm bandwidth, (c) 0.1 nm bandwidth, and (d) 1 nm bandwidth and spectral filters of 0.1 nm bandwidth applied to signal and idler, respectively.

Besides the reconstruction of spectral intensities, we can also use the information from the SFG measurements to derive the information about the JSA of the type-II SPDC photon-pair source for various experimental conditions. This is done by convoluting the joint spectral distribution obtained from the classical SFG measurements (Fig. 4(a)) and the spectrum of the pump, which is assumed to be a Gaussian with a center wavelength of 770.9 nm and a linewidth of 1 nm and is plotted in Fig. 7(a). The obtained JSA of the source near degeneracy is shown in Fig. 7(b), exhibiting a typical elliptic distribution [9] with major and minor axes along the directions defined by $\omega_s+\omega_i=\omega_p$ and $\omega_s=\omega_i$, where $\omega_p$, $\omega_s$, and $\omega_i$, are the angular frequencies of pump, signal, idler modes, respectively. The ratio $w_r$ of the widths of the intensity distribution along the two axes determines the spectral correlation (degree of nonseparability) between the two emitted photons of crossed polarization. The ratio $w_r$ is a function of the spectral widths of the pump, signal, and idler photons [30]. Fig. 7(c) shows the calculated JSA of the source obtained when the linewidth of the 770.9 nm pump is reduced to 0.1 nm and Fig. 7(d) depicts the JSA when the pump has a bandwidth of 1 nm but we assume spectral filtering of the signal and idler beams with filters centered at the degeneracy wavelength and with a bandwidth of 0.1 nm. The two means of spectral control on the pump and photon pairs produce an opposite effect on the spectral correlation property of the source, as expected. For applications using heralded single photons, the degree of factorability or the purity of the type-II photon-pair source can be elevated via performing spectral filtering to the photon pairs. The inverse Schmidt number or the purity $P=\sqrt{1-(w_r^2-1)^2/(w_r^2+1)^2}$ [9] has been effectively increased to ~99.86% for the joint spectral result obtained in Fig. 7(d), where the widths of the major and minor axes of the 2D distribution are ~0.25 and ~0.237 nm, respectively. In general, the purity also depends on the phase of the JSA, which we cannot measure. Hence, our estimate is an upper limit for the purity, although in the strongly filtered case we expect that the phase is nearly constant across the filter bandwidth, given that this is much narrower than the phasematching bandwidth. We note, that techniques based on stimulated emission or SFG can be also used to characterize the phase correlations in a photon-pair source [18, 20]. We conclude, that based on the measured SFG powers, the spectral properties of the photon-pair source for a variety of conditions can be quickly estimated. This facilitates the on-demand adaptation of these properties to dynamically changing demands.

Up to now, we discussed SFG and the prediction of the JSA of the two-photon source for the case that only fundamental waveguide modes are involved in the SPDC. However, higher-order waveguide modes may in general have different dispersion properties than fundamental modes, enabling the generation of different quantum states. Hence, characterizing the spectral properties of generated quantum states involving higher-order modes is important to fully use the potential of integrated photon-pair generation. The waveguides of our SPDC source support only fundamental modes for signal and idler in the telecom S, C, and L bands. However, in the spectral range of the pump at 725-825 nm wavelength, such waveguides support several modes [31]. We observed SFG to a higher-order mode in a waveguide with 9.6-µm poling period at 220 °C device temperature. Fig. 8(a) shows the measured spectral dependence of the power of the pump mode generated by SFG. In the accessible spectral range of this measurement, we observed one brighter spectral band coming from the more efficient SFG to the fundamental pump mode and one fainter band due to the conversion to a higher-order mode of the pump. The signatures stemming from the different pump modes are well separated, indicating that in SFG it is easily possible to independently assess signals belonging to either interaction. This may not be easily possible in SPDC, as clean excitation of higher-order modes in waveguides is challenging and measured SPDC spectra could show mixed signals stemming from a superposition of different pump modes. In Fig. 8(b), we finally plot the reconstructed JSA of a photon pair generated by SPDC from the higher-order pump mode to the H-and V-polarized signal and idler fundamental modes. Here, we assumed a pump centered at 770.9 nm with a bandwidth of 0.1 nm. Due to the weak dispersion of our waveguide, this JSA does not display markedly different properties compared to the ones involving only fundamental modes. Nevertheless, this result demonstrates the potential of the quantum-classical correspondence to also reconstruct quantum states generated involving higher-order modes.

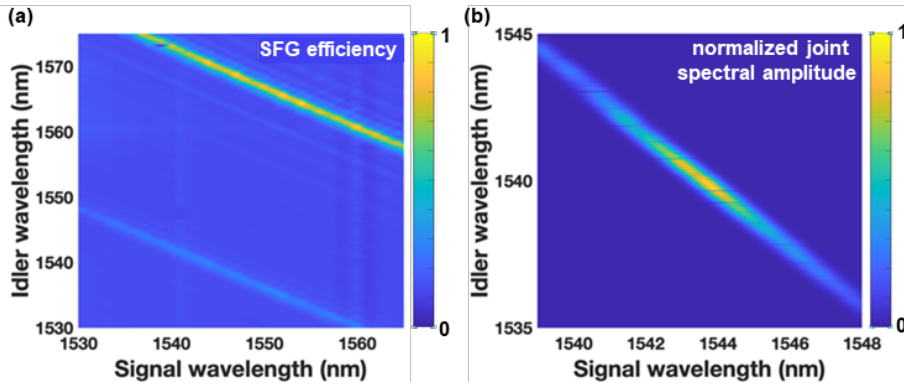

Fig. 8. (a) Intensity distribution of the SFG measured from the SPDC waveguide source with 9.6-µm poling period at 220 °C as a function of the signal and idler wavelengths. (b) JSA of the type-II SPDC for higher-order pump, calculated from the classical SFG measurement when assuming a pump at 770.9 nm with 0.1 nm bandwidth.

The results obtained in this work manifest that the use of classical SFG measurements can be a viable approach to fast and efficiently characterize the spectral information of a SPDC source based on a narrowband type-II phase-matching process. The approach makes possible the dynamic monitoring and stabilization of the spectra of (cross-polarized) photon pairs via feedback control mechanisms. Therefore, it can facilitate the realization of stable and high-purity photon-pair sources, which could be highly attractive to quantum communication applications. For example, in QKD systems using entanglement-based protocols [32], a highly stable entangled photon source is the key to effectively reduce the error rate and thus improve the key rate of the quantum transmission for increasing the practicability of the systems.

## 4. Conclusions

We successfully reconstructed the spectral properties of a type-II Ti:PPLN SPDC waveguide source based on the measurements of the SFG joint spectral efficiencies and the use of the quantum-classical correspondence describing the relationship between SFG and SPDC. In the study, we also performed direct spectral measurements of the generated cross-polarized photon pairs and found an excellent agreement of the results obtained from both the direct SPDC and classical SFG characterizations. The important equivalence between the quantum and classical conversion processes allows us to use a conventional SFG measurement technique to retrieve and monitor the quantum spectral performance of (cross-polarized) photon-pair sources dynamically, benefitting the purification and stabilization of the output quantum state of the sources, which could be of great interest to many applications including quantum communications. The reconstructed spectral information from the classical measurements indicates that the purity of our SPDC source can be ~99.86% at the degenerate wavelength when a 770.9-nm pump of 1-nm linewidth and spectral filters of 0.1-nm bandwidth for signal and idler modes are used. We also found that characterization with the SFG, an inverse process of SPDC, enables separation of the spectral properties of the photon-pair source stemming from interaction with different modes, e.g. different pump modes. This separation is complicated in the SPDC process, as excitation of a specific mode in a multimode waveguide is not trivial.

## Funding

Ministry of Science and Technology (MOST) of Taiwan (106-2221-E-008-068-MY3, 108-2627-E-008-001, and 109-2911-I-008-503); German Research Foundation (SE 2749/1-1, PE 1524/13-1); German Academic Exchange Service (ID 57448581); German Federal Ministry of Education and Research (FKZ 13N14877); Carl Zeiss Foundation; Australian Research Council (ARC) (DP190100277); Australia-Germany Joint Research Cooperation Scheme; Erasmus+ Mobility Program.## References